\documentclass[3p,preprint,authoryear]{elsarticle}
\usepackage{amsmath}
\usepackage{amssymb}
\usepackage{multirow}
\usepackage{gensymb}
\usepackage{graphicx}
\usepackage{float}
\usepackage{placeins}
\usepackage{epstopdf}
\usepackage{rotating}
\DeclareGraphicsExtensions{.jpg}
\usepackage[margin=0.2cm]{caption}

\begin{document}
\title{Characterizing Asteroids Multiply-Observed at Infrared Wavelengths}
%\author{Seth~C.~Koren} 
%\affil{Department of Physics \& Astronomy, University of Pennsylvania, %Philadelphia, PA 19104}
%\email{korens@physics.upenn.edu}
%\author{Edward~L.~Wright}
%\affil{UCLA Astronomy, PO Box 951547, Los Angeles, CA 90095-1547}
%\author{A.~Mainzer}
%\affil{Jet Propulsion Laboratory, Pasadena, CA 91109}
%\author{Carrie~R.~Nugent}
%\affil{Jet Propulsion Laboratory, Pasadena, CA 91109}

\author[UPenn]{Seth C. Koren\corref{mycorrespondingauthor}}
\cortext[mycorrespondingauthor]{Corresponding Author}
\ead{korens@physics.upenn.edu}

\author[UCLA]{Edward L. Wright}
\ead{wright@astro.ucla.edu}
\author[JPL]{A. Mainzer}
\ead{Amy.Mainzer@jpl.nasa.gov}
%\author[JPL]{Carrie R. Nugent}
%\ead{Carolyn.R.Nugent@jpl.nasa.gov}

\address[UPenn]{Department of Physics \& Astronomy, University of Pennsylvania, Philadelphia, PA 19104, USA, 215-898-8141}
\address[UCLA]{UCLA Astronomy, PO Box 951547, Los Angeles, CA 90095-1547, USA}
\address[JPL]{Jet Propulsion Laboratory, Pasadena, CA 91109, USA}

\begin{abstract}
We report Markov chain Monte Carlo fits of the thermophysical model of \citet{wright} to the fluxes of 10 asteroids which have been observed by both WISE and NEOWISE. This model is especially useful when one has observations of an asteroid at multiple epochs, as it takes advantage of the views of different local times and latitudes to determine the spin axis and the thermal parameter. Many of the asteroids NEOWISE observes will have already been imaged by WISE, so this proof of concept shows there is an opportunity to use a rotating cratered thermophysical model to determine surface thermal properties of a large number of asteroids.
\end{abstract}


\maketitle

\section{Introduction}

Thermophysical asteroid models have been in use for decades, and they have gradually improved in their completeness. The effects of thermal inertia in causing a time delay from local noon in the temperature maximum have long been taken into account \citep{peterson}.  The peaking of emission near zero observational phase angle (`beaming') was accounted for later in the Standard Thermal Model (STM) by calculating the emission at zero phase and then applying a linear correction factor for other phases. A beaming correction factor was then applied to account for the fact that beaming reduces reradiated energy; the STM is from \citet{lebofsky}. The beaming correction had also been used in \citet{jones}, and \citet{morrison}.  \citet{harris} improved the effectiveness of the STM with his Near Earth Asteroid Thermal Model (NEATM) by letting this correction factor be a free parameter, and adjusting it to match the observed color temperature when multiple thermally-dominated wavelengths are available. However, both of these are empirical models that account for a variety of phenomena and parameters using only the beaming parameter.

\citet{hansen} did not use a beaming model and instead considered an asteroid as covered in craters, so that at non-zero phase angle there is increased shadowing of the visible portion of the asteroid over what would be observed from a smooth surface. This dampening of flux at non-zero phase can be interpreted as a peak at zero phase. \citet{spencer} adds to this model the consideration of light reflecting off different parts of craters, as well as an iterative numerical process to model heat conduction. \citet{lagerros} combines the effects of both thermal inertia and cratering in his model, and calculates a correction factor based on a comparison between a smooth surface and one with craters, though a more detailed discussion as to the effects of different sorts of surface roughness is given in \citet{lagerros2}. \citet{delbo} includes surface roughness by considering the mean slope of the surface of the asteroid, rather than assuming any sort of crater geometry. \citet{hanus} uses optical photometric data to investigate shape models of the asteroid before applying a thermophysical model using infrared data.

\citet{wright} takes the surface cratering into account explicitly in his Spherical, Cratered, Rotating, Energy-conserving Asteroid Model (SCREAM, name assigned here for ease of reference) by including the local effects of this geometry in the power balance calculations of the temperature distribution over the surface of the asteroid. As a result energy is entirely conserved, and the model can include the effects of the reflection of solar light and the absorption of blackbody radiation caused by the mutual visibility of different parts of a crater, in addition to considering vertical heat conduction. Other asteroid thermophysical models that take all of these into account include those of \citet{muller}, \citet{rozitis}, and \citet{leyrat}.

The Wide-field Infrared Survey Explorer (WISE) mission has provided a veritable treasure trove of information on the infrared sky \citep{WISE}. This includes asteroids, of which over 160,000 have now been observed. The NEOWISE project allowed individual exposures from WISE to be publicly archived and searched for moving objects, to enable the discovery of new asteroids and comets \citep{mainzera}.

The NEATM makes it possible to quickly perform thermal modeling of asteroids and has already been used on WISE data \citep[e.g.][]{mainzerc}. While the SCREAM is much more computationally intensive, it has the potential to allow additional parameters beyond diameter, albedo, and beaming to be determined, such as spin axis and thermal inertia. Parameters such as thermal inertia and spin axis can be more narrowly constrained when observations of an asteroid are available at multiple epochs. In cases when multiple viewing geometries are available, the NEATM can converge to different beaming factors at each epoch (though this can also result just from asphericity or different viewing geometries). Multiple epochs of observation are very advantageous in the SCREAM, as the differing phase angle gives views of different local times and/or latitudes of the asteroid, which allows one to characterize the asteroid spin axis in order to explain the phase-varying flux. 

With the recent reactivation of the WISE telescope for the restarted NEOWISE mission, many asteroids are now being reobserved at different phase angles \citep{mainzer4}. This new mission thus gives us an opportunity to characterize these asteroids using the SCREAM, with which we can jointly fit all the data to explain the phase-varying flux. As a proof of concept, we here report Markov chain Monte Carlo fits of the SCREAM to 10 asteroids which have already been reimaged by the NEOWISE mission. \\

\section{Data}

Candidates for analysis were found by querying the Minor Planet Center\footnote{http://www.minorplanetcenter.net/} (MPC) for all WISE and NEOWISE observations of asteroids, and then searching through the output for asteroids which were seen by both. Then the Infrared Science Archive's\footnote{https://irsa.ipac.caltech.edu/frontpage/} moving object search feature was used to find flux data for each of the asteroids. After throwing away temporal outliers ($>1$ day from other observations), the data were binned into time series from different observational epochs, and then the interquartile mean (or mid-average) of each epoch was taken as the new data point. Since our asteroids all have prograde orbits with periods $\gtrsim 1$ yr, no meaningful intra-bin trends were seen in the observational epochs, which were of length $\lesssim 10$ days. A new uncertainty for each data point was calculated as:

\begin{equation}
\sigma_{f,i} = \sqrt{\frac{C^2}{N/2}\sum\limits_{j = N/4 + 1}^{3N/4} \sigma_j^2 + \left(\frac{0.1 \ln(10)}{2.5} f_i\right)^2}
\end{equation}

\noindent where there are $N$ sorted observations being mid-averaged with uncertainty $\sigma_j$, and their mid-average flux value is $f_i$. The first term is the standard combination of independent uncertainties applied to the second and third quartiles of the data, but the correction factor $C \simeq 0.77$ accounts for the extra information from the data points we discarded and was found via Gaussian error modeling\footnote{Pseudocode: \\ \indent \indent do M times \\ \indent\indent\indent take N samples from a unit normal distribution \\ \indent\indent\indent calculate the standard deviation of the combined second and third quartiles \\ \indent\indent calculate the average standard deviation \\ \indent\indent divide by the standard deviation of N/2 samples from a unit normal distribution of $1/\sqrt{N/2}$ \\ This calculation produces $C \simeq 0.77$, indicating that the data in the first and fourth quadrants which are thrown out are nonetheless adding information to our statistic and so need to be accounted for. }. The second term is the equivalent of 0.1 magnitudes, and was added to account for the magnitude of the approximations made in our model which are detailed in Section \ref{methods}, especially the discretization of the craters.

A summary of the data used can be found in Table \ref{data}.

When many high-accuracy observations of an asteroid over its rotational period are available, one may use a technique called `lightcurve inversion' to deduce both the shape and rotational characteristics of the asteroid \citep{kaasalainen}. However, in our relatively low S/N regime this method is not so useful. By binning our data over entire observational epochs, we average over the periodic flux variations due to the asphericity of the asteroid and solve for an `effective diameter'. Our interquartile mean provides us with statistically robust data at each viewing geometry. As a test, we did perform fits for 2 of our asteroids using all the observations separately, without binning, and the results were found to agree well with the results found using our binning process. For more on the assumption of sphericity in our model, see Section \ref{discussion}.

In our modeling we used Keplerian orbital parameters from the MPC, and found the absolute magnitudes of the asteroids using the JPL Horizons web interface\footnote{http://ssd.jpl.nasa.gov/?horizons} which were assigned an uncertainty of 0.3 magnitudes, as was done in \citet{mainzera} and \citet{mainzerc}. \\

\section{Methods}\label{methods}

Markov chain Monte Carlo (MCMC) methods sample probability distributions by constructing a Markov chain in state space which converges on the desired equilibrium distribution. A discussion of the mathematics behind the algorithm is beyond the scope of this paper \citep[see][]{mackay}, but the method is often used in astronomy to sample posterior probability distributions of free parameters in a model given some data. It is useful to think of a Markov chain as a biased random walk, where the bias is such that the `walker's' steps converge to the desired probability distribution. Here, this is accomplished by defining a likelihood function using the familiar $\chi^2$ statistic which has as its equilibrium distribution the likelihood of a given parameter vector $\bf\Xi$ being the `true' parameter vector. We define:

\begin{equation}
L[{\bf\Xi}] = \kappa e^{-\frac{1}{2}\chi^2[\bf\Xi]} 
\end{equation}
\begin{equation}
\chi^2 = \sum_{i} \left( \frac{f_{data,i} - f_{model,i}[{\bf\Xi},t_i]}{\sigma_{f,i}} \right)^2
\end{equation}

\noindent where $i$ indexes the data points, each of which has a flux $f_{data,i}$, an uncertainty on that flux $\sigma_{f,i}$, and a time of observation $t_i$. $\kappa$ is a normalization constant which may be ignored for our purposes since MCMC methods evaluate only $ {L[{\bf\Xi}_1]}/{L[{\bf\Xi}_2]} $ to determine the acceptance or rejection of the next parameter vector. We have assumed a diagonal covariance matrix on the data in our $\chi^2$ equation for simplicity, which should be a good approximation. We used the $emcee$ package for our MCMC analysis \citep{foreman}. $emcee$ provides an `ensemble sampler' which is affine-invariant and utilizes a large number of `walkers' to efficiently explore and sample parameter space, while employing parallelization to reduce the computational time needed for sampling. Affine-invariance ensures that the performance of our MCMC is not affected by correlations between our parameters causing anisotropic probability distributions \citep{goodman}.

Our thermophysical model has five free parameters:
\begin{itemize}
\item $\varphi$ - The RA of the spin axis of the asteroid
\item $\theta$ - The Dec of the spin axis of the asteroid
\item $\Theta_1$ - The dimensionless thermal parameter of \citet{spencer0} computed at a distance of 1 AU
\item $\epsilon$ - The emissivity of the asteroid surface
\item $D$ - The effective spherical diameter of the asteroid
\end{itemize}

The thermal parameter is a measure of the importance of thermal inertia on the temperature. It is defined as $\Theta = \frac{\sqrt{\kappa \rho C \Omega}}{\epsilon \sigma T_{\circ}^3}$, where $\kappa$ is the thermal conductivity of the regolith, $\rho$ is the density, $C$ is the specific heat per unit mass, $\Omega$ is the rotational frequency of the asteroid, $\sigma$ is the Stephan-Boltzmann constant, and $T_{\circ}$ is the equilibrium temperature of a facet on the asteroid oriented toward the Sun. The combination $\Gamma = \sqrt{\kappa\rho C}$ is known as the thermal inertia \citep{winter}. The equilibrium temperature is $T_{\circ} = \left(\frac{(1 - A) L_\odot}{4\pi\epsilon\sigma r^2}\right)^{1/4}$, where A is the Bond albedo defined below, $L_\odot$ is the solar luminosity, and $r$ is the distance between the asteroid and the Sun. The parameter we fit is the thermal parameter when the asteroid is at a distance of 1 AU. For each data point we transform $\Theta_1$ as $\Theta(r) = (r/1 \text{AU})^{3/2} \Theta_1$, using the value of $r$ during that observation, in order to get the thermal parameter that should be used in calculating our model fluxes for that data point. We approximate the emissivity as independent of wavelength. Since we approximate the asteroid as spherical, the diameter we fit is an effective diameter.

At each MCMC step, we use the diameter and the absolute magnitude of the asteroid to calculate a Bond albedo $A$ for use in our calculations. The relation used is:
\begin{align*}
A &= q \left(\frac{1329 \ \text{km} \ 10^{-H/5}}{D}\right)^2 \\
q &= 2 \int_{0}^{\pi} \frac{I(\alpha)}{I(0)} \sin\alpha \ d\alpha
\end{align*} 

\noindent where $q$ is known as the `phase integral', and accounts for the fact that the brightness $I$ of the asteroid may depend upon the phase $\alpha$. $q$ may be evaluated analytically using a full model of the reflectivity of an object's surface, or an empirical number may be calculated. We use the $q = 0.384$ value used in \citet{mainzerb}, which is based on the IAU Commission 20 recommended value of the slope parameter for the magnitude-phase relationship of $0.15$ in the $H-G$ model of \citet{bowell0}. Since the true value of $q$ is not well known, this could be a possible cause behind cases where the model fails. An incorrect value of $q$ or $H$ could affect the probability distribution for the diameter in a complicated way, since $D$ enters our model both by itself and through $A$. Where we have data at all four wavelengths, we fit for the albedo in the two shortest wavelengths, making the simplifying assumption that the albedo at $3.4 \mu$m is the same as the albedo at $4.6 \mu$m \citep{mainzerb}.

%We assume that the albedo does not appreciably change between the $3.4 \mu m$ and the $4.6 \mu m$ bands, as in \citet{mainzerb}. We note that although the albedo of the asteroid may take different values in the near-infrared ($3.4 \mu m$ and $4.6 \mu m$) and mid-infrared ($12 \mu m$ and $22 \mu m$), the albedo used in the SCREAM is an average albedo over the bands weighted by the solar flux distribution, so we effectively find the near-infrared value. Thus we don't include two albedos as free parameters in our model, since the thermal contribution of mid-infrared light is vanishingly small from Planck's law for blackbody radiation and the temperature of the Sun. 

We used priors on the angular elements to enforce the modular definition of the spin axis, and used a uniform prior of $0.9 \leq \epsilon \leq 1.0$ to force a physically sensible solution \citep{salisbury}, since the emissivity is in general poorly constrained.

The computationally-heavy step of our MCMC is computing $f_{model,i}[{\bf\Xi},t_i]$. \citet{wright} gives a description of the SCREAM algorithm. Here, we present details of the specific numerical methods used in our pilot study. As a general note, a large amount of time can be saved by taking full advantage of array operations and linear algebra in the following calculations and manipulations.

Our primary concern is to find the temperatures of facets of representative craters at different latitudes on the surface of the asteroid over time by solving the power balance equation. For this we must calculate the sources and sinks of power for each facet of each crater. As \citet{wright} lays out, we account for the incident solar flux, the blackbody radiation of the facet, solar reflection and blackbody radiation from other visible facets, and vertical heat conduction into and out of the surface of the facet. 

First, from the Keplerian orbital elements of the asteroid and the time of the observation, the positions of the Sun and Earth are calculated in the asteroid-centric frame. We then go about creating the facets of our representative crater at a particular latitude and local time. We utilize a cylindrical projection to go from facets in the 2d Cartesian square $(x,y) \in [-1,1]\times [-1,1]$ to the 3d spherical cap geometry we require via the transformation:
\begin{equation*}
\theta = \frac{\theta_{max}}{\pi} \left[\arcsin(y) + \frac{\pi}{2}\right]
\end{equation*}
\begin{equation*}
\varphi = \pi x
\end{equation*}
\noindent where the radial size is of no consequence since these are only representative craters, and $\theta_{max} = \pi/4$ was used in our modeling. This transformation maps lines parallel to the $x$ axis to azimuthal circles around the spherical cap, and lines parallel to the $y$ axis to great circles through the zero polar angle point. It also makes it simple to ensure our facets have equal surface area through the correct choice of bounding boxes before projecting, which we can calculate by changing variables under our map. 

Chains with $\theta_{max}$ as an additional free parameter were run for several asteroids. Minimum $\chi^2$ went down only marginally in each case, and the posterior probability distributions were found to be wide with a peak near $\pi/4$. Thus $\theta_{max} = \pi/4$ was used to model the generalized surface roughness which causes the beaming effect. Other works have taken similar strategies of removing $\theta_{max}$ as a free parameter, with \citet{hansen} assuming $\theta_{max} = \pi/3$ and \citet{rozitis} performing fits using a few different angles, for example.

We must choose how fine our discretization of the spherical cap will be. We decide for simplicity's sake to use the same number $n$ of facets in both the polar and azimuthal directions, for a total of $n^2$ facets. Fully optimized, our code is $\sim O(n^2)$, so we have a trade-off between computational expense and model accuracy.

Slightly different temperature distributions are found for models with greater numbers of facets due to partially-shadowed facets on crater rims being broken into some shadowed facets and some lit facets. While the fraction of the facets which are lit is the same in each model, a model with more facets more accurately calculates the average surface normal vector of lit facets in the crater. Since this causes temperatures to change, it creates a linear dependence on frequency of the radiated flux change due to the exponential factor in Planck's law. We note that the fluxes do approach some limit as the facet number increases, and $32^2$ facets gets close. If we go down to a model with $12^2$ facets, the cumulative loss in accuracy is usually $\sim6\%$ at the high frequency end, which is well within the minimum $0.1\ln10/2.5\% \simeq 9\%$ uncertainty we have provided for our data points. Since this is a proof-of-concept study and our computational resources were limited, we performed our calculations with 144 facets.

It's important to note that in general the error from using fewer facets depends upon the phase angle of the observation and the parameters used in the modeling. Moreover, in some cases, this error can be higher than the typical $<9$\% uncertainty. After we found best-fit parameters for each of our asteroids, we rechecked the loss in accuracy with these parameters for the specific viewing geometry of each data point. In each case, the flux errors were less than or comparable to the uncertainty we added to our data points, and corresponded to changes in predicted brightness temperature for the asteroid of less than 1K for all frequencies. This level of accuracy was deemed sufficient for this proof-of-concept study.

Once we have chosen our facet division, we set up the surface area integral and set it equal to one facet area:

\begin{equation*}
A = \int_{\theta_j}^{\theta_{j+1}} \int_{\varphi_i}^{\varphi_{i+1}} r^2 \sin\theta \ d\theta \ d\phi = \frac{2\pi(1-\cos\theta_{max})}{144}
\end{equation*}

We now change variables under the aforementioned map taking $r = 1$ for ease of evaluation, and we require $\left\{ x_i,y_j \right\}_{i,j = 0..12}$ to satisfy the following condition:

\begin{align*}
\frac{(1-\cos\theta_{max})}{144} &= (x_{i+1} - x_i) \\ &\times \sin\left[\frac{\theta_{\max}}{2\pi}\left(\arcsin(y_j) + \arcsin(y_{j+1}) + \pi\right)\right] \\ &\times \sin\left[\frac{\theta_{\max}}{2\pi}\left(\arcsin(y_{j+1}) - \arcsin(y_{j})\right)\right]
\end{align*}

We take $x_i = \frac{2\pi}{12} i$ for azimuthal symmetry, and then can easily solve for the $\left\{y_j\right\}$ using computational methods.

Defining the facets in this manner places a spherical cap at the origin opening downward, so we must maneuver it into the correct position.  We first place it on the midnight longitude at the point on the surface of the asteroid directly opposite from the Sun, and then rotate it to the equator around the cross product of the spin axis and the location of the Sun. Then we can simply rotate it to the right latitude, and then around the spin axis to the correct local time.

\begin{figure}[t]
\centering
\includegraphics[width=0.5\columnwidth]{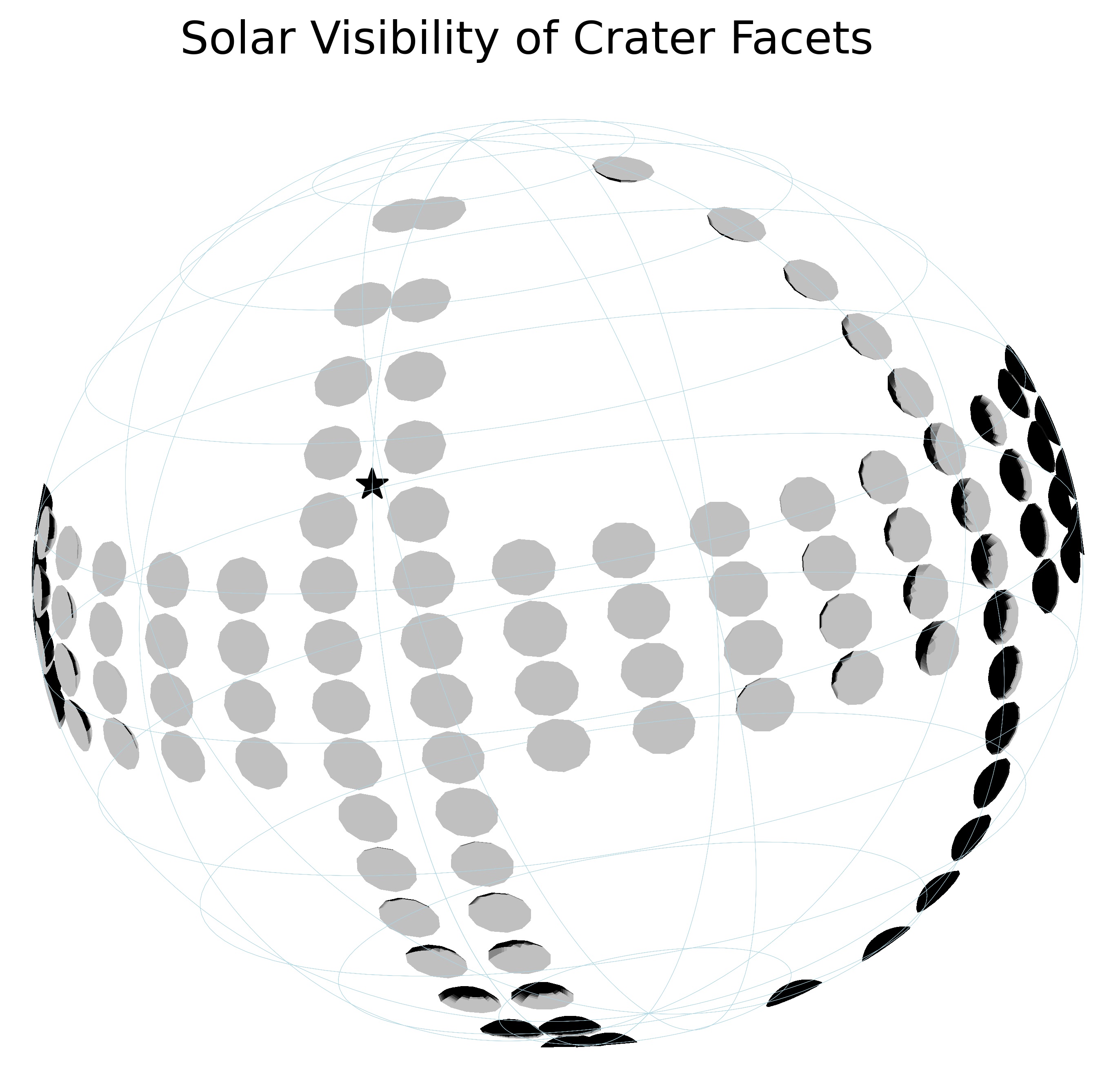}
\caption{A visualization of the solar visibility of some representative craters over the face of a spherical asteroid. The lighter parts have a direct line of sight to the Sun, and the dark parts are in shadow. The sub-solar point is marked with a star. \label{SolarVisibility}}
\end{figure}

With the facets in place, we can compute the cosines of the angles between the surface normals and the Earth and Sun. We can now also compute the `solar visibility fraction' of each facet - that is, the fraction of the facet which has a direct line of sight to the Sun, and thus incident solar flux. See Figure \ref{SolarVisibility} for an idea of the geometry involved.

This is accomplished by considering a finer breakdown of each facet into 100 `pixels', and calculating the point of intersection between the sphere of the crater and the line through each pixel and the Sun. For a sphere defined as $\left|\bf x - c\right|^2 = 1^2$ with $\bf x$ the points on the sphere, $\bf c$ the center, and the radius taken to be 1; and a line defined as ${\bf x} = {\bf o} + s {\hat{\ell}}$ with $\bf o$ the pixel location, ${\hat{\bf \ell}}$ the unit vector between the pixel and the Sun, and $s$ a free variable over the reals, we can calculate the distance $s$ from the pixel at which the intersection lies as $s = 2{\hat{\ell}}\cdot (\bf c - o)$. If the point given by ${\bf o} + s {\hat{\ell}}$ is farther from the Sun than the pixel in question (if $s < 0$) or the point is closer to the bottom of the crater than $\sqrt{1 + \sin^2\theta_{max}}$, the maximum extension for our spherical cap, then the ray between the pixel and the Sun does intersect the crater, so the pixel does not see the Sun.  If not, then there will be solar flux incident upon that pixel. We do this for all of the 100 pixels in each facet, and then can calculate the total solar power incident upon the facet.

Once we have calculated the solar visibility fractions of each facet of a crater at a given latitude at each time step as it rotates around the asteroid, we must determine the vertical heat conduction before we can solve the power balance equation.

The vertical heat conduction coefficients $G$ couple the temperature of a facet at one time to the temperature of the same facet at different times. We calculate them by computing the sum given in \citet{wright} Equation 14. Due to the $n^{-2.5}$ dependence here, we can cut off the sum after 100 terms with little loss ($<$ 1.2 \%), which speeds up this step a fair amount. Once this step is complete, we have precomputed the coupling coefficients between the temperatures of all the facets of a crater at all times. We then need to solve this system of equations simultaneously for the temperatures of the facets.

To do so, we create a function that takes in $32 \times 144 = 4608$ temperatures of crater facets at a given latitude over time, and returns the residuals of the power balance equation calculated by subtracting the RHS of Equation 21 in \citet{wright} from the LHS. Even though the equations are coupled and non-linear, they can still be solved using iterative numerical methods. The best method found was a Newton-Krylov solver implemented in the \textit{SciPy} Python package\footnote{http://docs.scipy.org/doc/scipy/reference/generated/scipy.optimize.newton\textunderscore krylov.html}, which uses the inverse Jacobian to iteratively minimize the residuals. This solver generally returns sensible solutions, and converges relatively quickly to the $\max(\text{residuals}) < 10^{-3}$ level in $\sim100$ evaluations. Examples of the temperature distribution returned by the solver can be found in \citet{wright}.

After running this for each latitude, we can use the temperatures of each facet of each crater over the asteroid to calculate the total blackbody radiation emitted by considering each as an independent blackbody. We add to this the reflected flux from the Sun by multiplying the solar flux by the appropriate geometric factors we found earlier.

While we can evaluate this predicted flux at any frequency, rather than integrating over WISE's passband spectral responses ourselves we use the linear quadrature formulae given in \citet{lambda}. We compute the predicted flux at 2 or 3 frequencies per passband, and by summing these with the appropriate weights we obtain analogues for WISE flux observations which are accurate to within half a percent.

Now we can calculate the $\chi^2$ statistic by comparing these predictions to the data, and then feed that to the MCMC. Theoretically, we could start the Markov chain Monte Carlo at any point in parameter space, let it run, and - once we've waited long enough - return to find that the output of the chain has converged exactly to the posterior probability distributions we want. However, between the large number of computations needed to evaluate the model and the dimensionality of our parameter space, this approach is computationally unfeasible. We instead assist $emcee$ in finding the global minimum, and then use it to map out that minimum and give us the posterior probability distribution.

We begin by finding a reasonable range for the asteroid diameter by calculating the $\chi^2$ with various diameter guesses and by visual comparison with the overall magnitude of the flux. This process could easily be automated by testing some set values of the other four parameters with a set of diameter values, and is only done to save computational time. Then we coarsely sample parameter space with $emcee$ by running a chain beginning at a widely-distributed `sample ball', which starts walkers covering essentially the entire parameter space in $\varphi, \theta, \epsilon$ and $\Theta_1$, as well as the conceivable range in $D$. After we are satisfied that the parameter space has been well-investigated and decent $\chi^2$ wells have been found, we restart the chain centered on these different wells with smaller walker balls to find the minima of the wells. After investigating any promising wells, we are reasonably satisfied that the deepest of them is the global minimum. For our final sampling, we start two chains at this minimum of 50 steps with 50 walkers each, and then throw away the first 10 steps from each chain to account for `burn-in'.

\section{Results}\label{results}

Characterizations of the posterior probability distributions (PPDs) of the parameters for each of the asteroids can be found in Table \ref{ppd}. Estimates of the thermal inertias of the asteroids can be found in Table \ref{thermal}. Only fits where the best $\chi^2$ per degree of freedom was less than $1.5$ are included in these tables. The number of degrees of freedom is the number of observational epochs multiplied by the number of operational passbands in each observational epoch, less the five fit parameters in our model. Ten out of the twenty asteroid fits which were attempted yielded such fits.

%The emissivities are in general not very well characterized, usually turning out rather widely distributed over our prior. This ends up increasing the uncertainty on the diameter, as the two parameters are partially degenerate \citep[see][Equation 22]{wright}. Removing the emissivity as a free parameter and using some specific guess value would lower this uncertainty but is not physically justified, as we should expect asteroids to have varying emissivities.

\begin{figure*}[h]
\centering
\includegraphics[width=\textwidth]{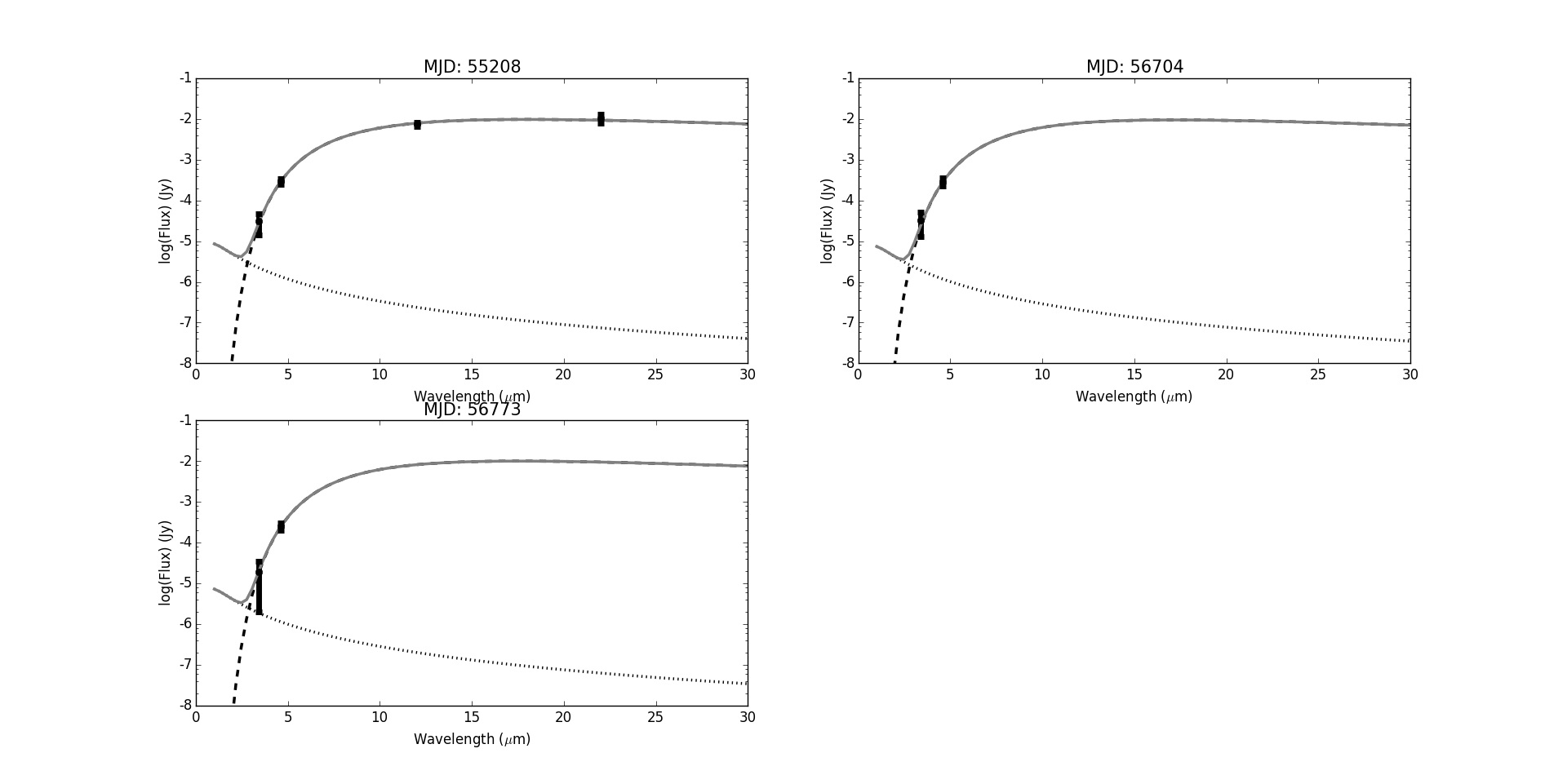}
\caption{Best fit solution for the asteroid 2014 HJ129 with 3 DOF and $\chi^2 = 0.60$. Blackbody radiation is the dashed black line, reflected solar light is the dotted black line, and their sum is the solid grey line. Data points are plotted with error bars, which may be too small to be seen. The fluxes are not normalized for different heliocentric distances between epochs, but appear as observed. \label{2014HJ129Best}}
\end{figure*}

\begin{figure*}[h]
\centering
\includegraphics[width=\textwidth]{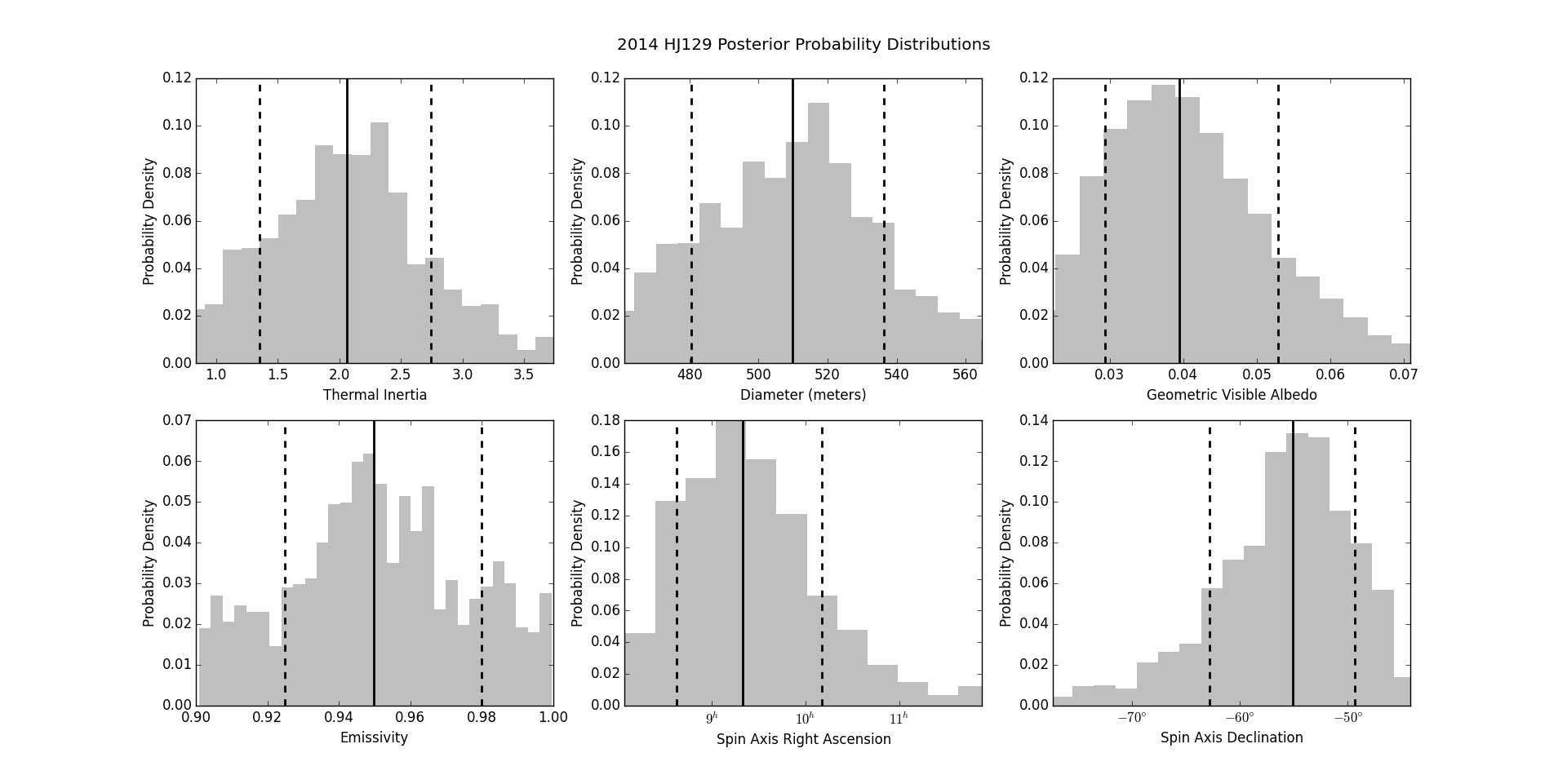}
\caption{Posterior Probability Distributions from a Markov chain Monte Carlo fit of the SCREAM applied to the asteroid 2014 HJ129. Visible on the plots are the 2.5\% to the 97.5\% bins, except for the emissivity in which the entire allowable range is shown. The solid vertical line denotes the median value, and the dashed vertical lines denote the $\pm 1 \sigma$ values, as reported in Table \ref{ppd}. \label{2014HJ129PPDs}}
\end{figure*}

\begin{figure*}[h]
\centering
\includegraphics[width=\textwidth]{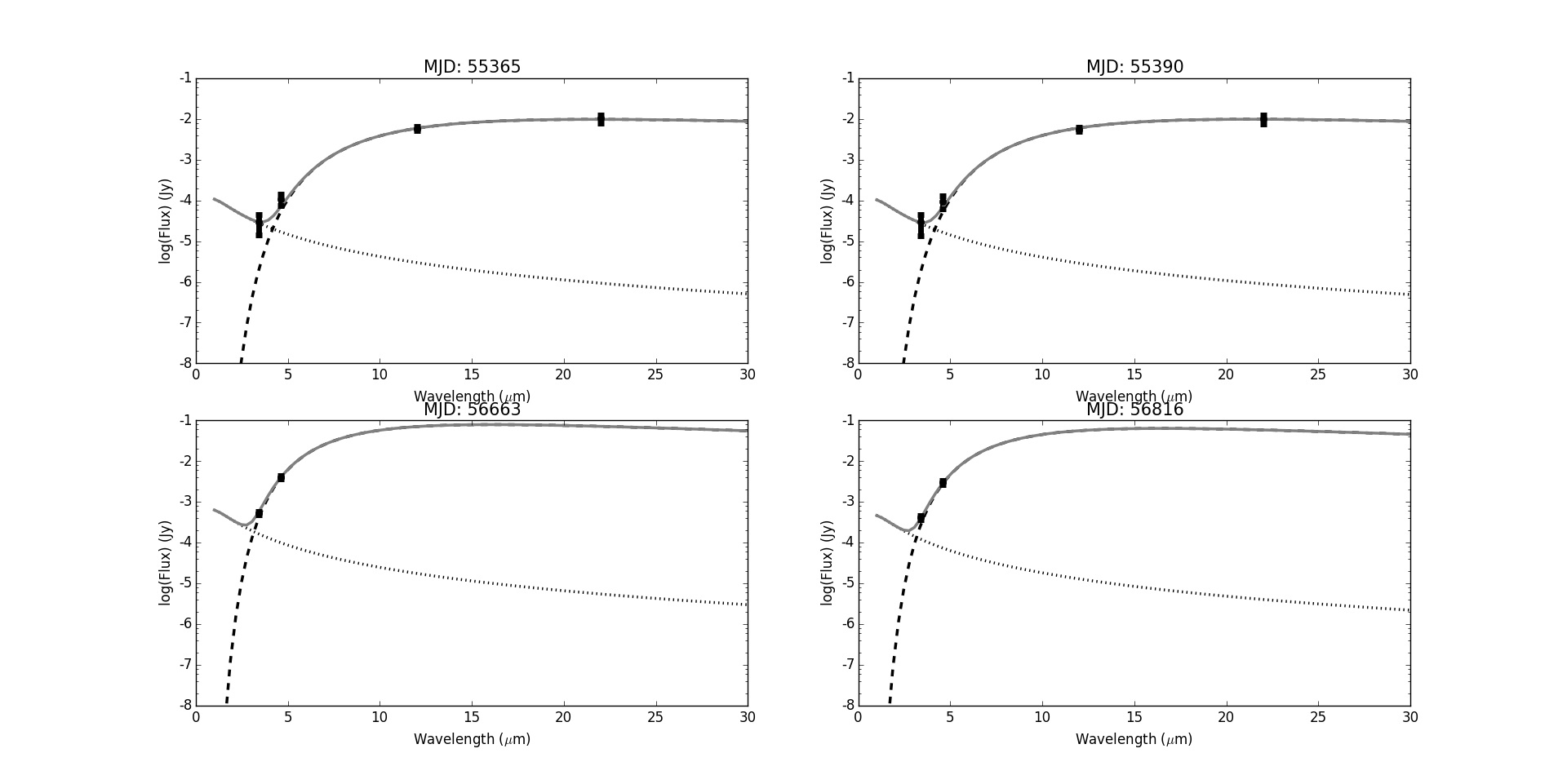}
\caption{Best fit solution for the asteroid 2102 Tantalus with 7 DOF and $\chi^2 = 2.84$. Blackbody radiation is the dashed black line, reflected solar light is the dotted black line, and their sum is the solid grey line. Data points are plotted with error bars, which may be too small to be seen. The fluxes are not normalized for different heliocentric distances between epochs, but appear as observed. \label{1975YABest}}
\end{figure*}

\begin{figure*}[h]
\centering
\includegraphics[width=\textwidth]{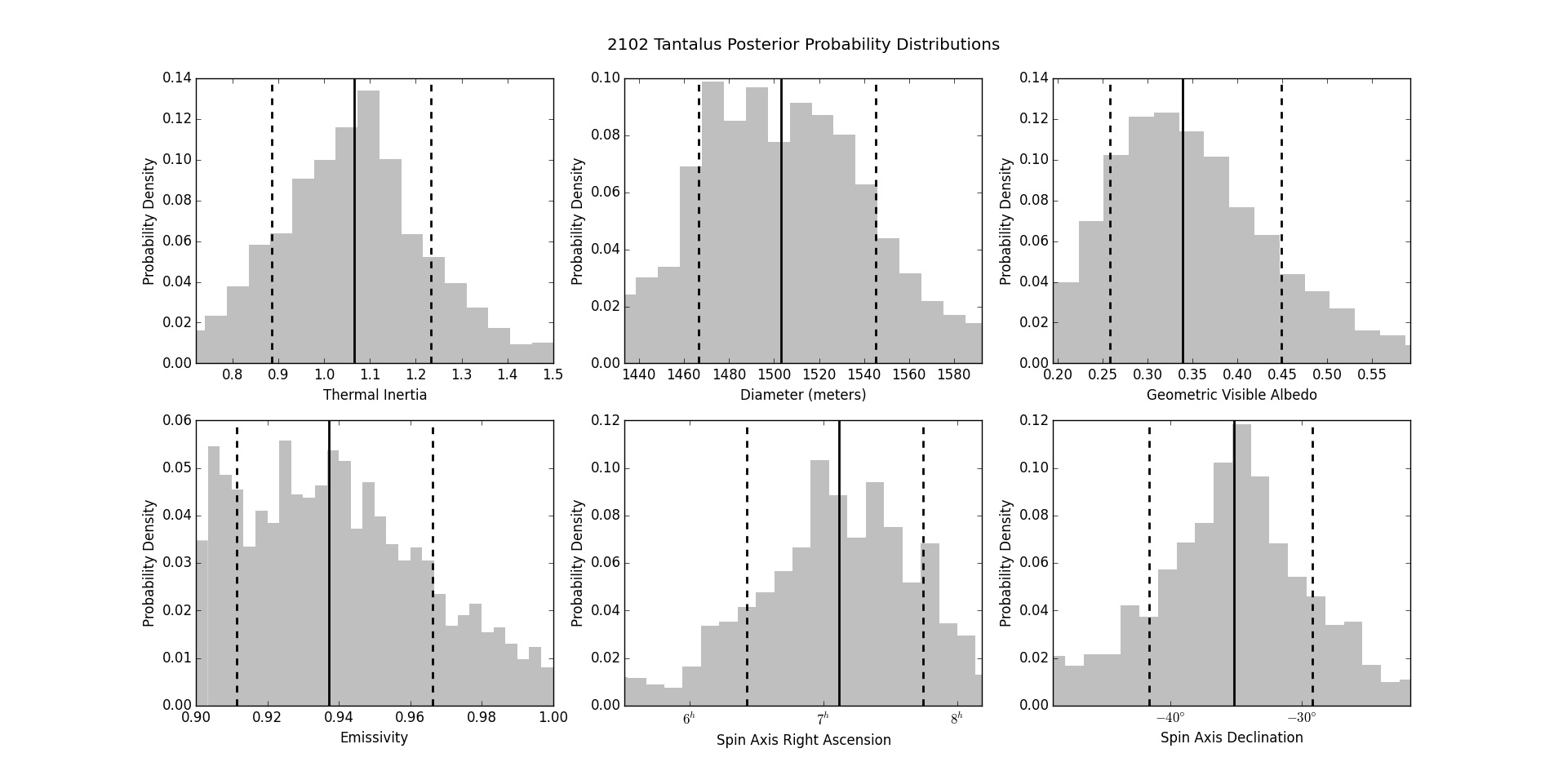}
\caption{Posterior Probability Distributions from a Markov chain Monte Carlo fit of the SCREAM applied to the asteroid 2102 Tantalus. Visible on the plots are the 2.5\% to the 97.5\% bins, except for the emissivity in which the entire allowable range is shown. The solid vertical line denotes the median value, and the dashed vertical lines denote the $\pm 1 \sigma$ values, as reported in Table \ref{ppd}. \label{1975YAPPDs}}
\end{figure*}

For some examples of good fits, the best fit solutions for the asteroids 2014 HJ129 and 2102 Tantalus are plotted in Figures \ref{2014HJ129Best} and \ref{1975YABest}, respectively. The PPDs returned by the MCMC can be seen in Figures \ref{2014HJ129PPDs} and \ref{1975YAPPDs}. 

\begin{figure*}[h]
\centering
\includegraphics[width=\textwidth]{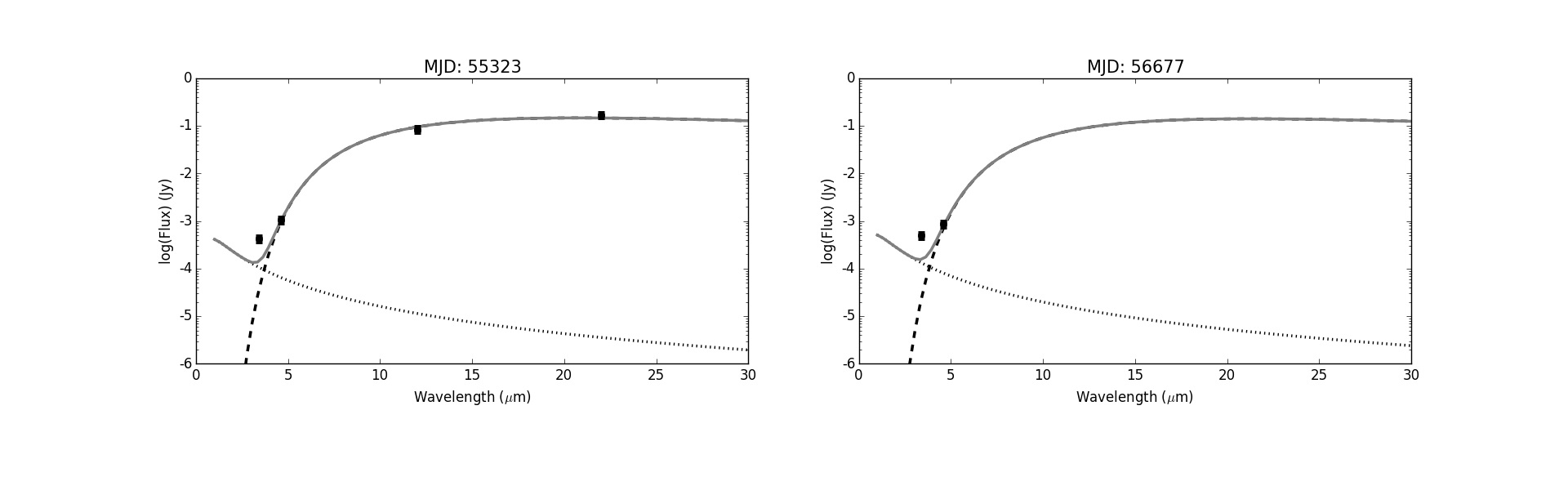}
\caption{Best fit solution of the failed fit for the asteroid 1627 Ivar with 1 DOF and $\chi^2 = 90.76$. Blackbody radiation is the dashed black line, reflected solar light is the dotted black line, and their sum is the solid grey line. Data points are plotted with error bars, which may be too small to be seen. The fluxes are not normalized for different heliocentric distances between epochs, but appear as observed.\label{1929SHBest}}
\end{figure*}

For an example of a failed fit, the best fit solution found for the asteroid 1627 Ivar is plotted in Figure \ref{1929SHBest}. While it is apparent from the figure and the best $\chi^2$ per degree of freedom found of $90.76$ that the SCREAM does not model this asteroid well, we are unable to say anything concrete about why the SCREAM failed in specific cases. All that is apparent is that no set of asteroid geometric and thermal properties could be found which produced the correct flux curves that 1627 Ivar was observed to have. However, this asteroid is one of those \citet{hanus} found a good fit for using their varied shape thermophysical model, indicating that it is likely that the assumption of sphericity is behind the failure of the SCREAM in this case.

%Two dimensional histograms of the RA and Dec PPDs show the location of the spin axis on the sky in Figures \ref{2014HJ129JointHist} and \ref{1975YAJointHist}.

\section{Discussion}\label{discussion}

Since the model makes a number of assumptions, it is not surprising that good fits were not found for all asteroids. No beaming effect takes place with a sufficiently smooth asteroid, so a less-cratered model would be more appropriate in that case, and could perhaps be addressed by including a crater-covered fraction parameter in our model. For slowly-rotating asteroids, the data which were binned in each observational epoch may not have covered an entire period, or may not have been uniformly distributed over the asteroid's rotation. This could cause our flux interquartile means to poorly represent the effective spherical diameter. Asteroids which are very aspherical could have a similar problem. We note that the SCREAM can be used to calculate fluxes for asteroids of arbitrary convex shapes \citep{wright}. While this generalization is useful, convexity is not always a good assumption, especially in the case of bifurcated asteroids (near 10\% of radar-imaged near Earth asteroids larger than $\sim$ 200m are bifurcated candidate contact binaries \citep{benner}) and exotic shapes like that of 25143 Itokawa \citep{demura}. A new method would be needed to model more complicated shapes. Attempting fits of aspherical convex shapes to asteroids for which the spherical model failed should be done in follow-up work, as should an extension of this model to include a parameter like a crater-covered fraction. The use of the SCREAM in the varied shape thermophysical model method of \citet{hanus} when optical photometric data are available may also provide a fruitful way of incorporating shape effects.

After obtaining our results, we examined the characteristics of our asteroids as well as our data for trends which might explain for which asteroids good fits could not be found. No patterns in orbital characteristics were found. No good fits could be found for any of the four asteroids with absolute magnitude H brighter than 15.0. Additionally, the SCREAM failed for four of the five asteroids with the largest normalized lightcurve amplitudes \footnote{Normalized lightcurve amplitudes were calculated for each observing epoch and each band by sorting the data points, subtracting the 3N/4th value from the N/4th value - where N is the number of observations in that epoch - and dividing by the interquartile-averaged uncertainty of that epoch. We then averaged this measure over all of the observing epochs and bands for each asteroid.}. Overall the average normalized lightcurve amplitude of the asteroids for which good fits were not found was 3.2, while for the asteroids for which good fits were found the average was 2.8. Together, these suggest that shape effects are the largest problem for the SCREAM, which is unsurprising. Note that the lightcurve amplitude may be underestimated in the case of slowly-rotating asteroids, so it's possible that some of the other failures are due to asteroids with significant shape effects which are rotating slowly enough for our method of lightcurve amplitude evaluation to be unable to discern.

We also compared our modeled diameters and albedos to fits of the NEATM to our asteroids which have been previously published by the NEOWISE team in \citet{mainzerc} and \citet{mainzer4}. See Table \ref{comparison} for a comparison of the two. The two measures are generally in good agreement.

Two of the asteroids we fit have classified taxonomic types: 2102 Tantalus is Q type and (89355) 2001 VS78 is Sr type, placing both in the roughly stony category \citep{neese}. The relatively high albedos found for both of these asteroids are consistent with this classification. 

Having a further empirical method for evaluating the fit parameters would allow us to check the external consistency of the SCREAM and our characterization methods. Unfortunately, radar data do not exist for any of the asteroids we characterized\footnote{http://echo.jpl.nasa.gov/asteroids/index.html}. A follow up comparison once such data are available would be prudent. \\

\section{Conclusion}

The SCREAM can be used to derive asteroid effective spherical diameter, emissivity, albedo, thermal parameter, and spin axis when multi-epoch thermal IR data are available. Future follow-up work is needed to determine the accuracy of our fits. 

The information this gives us about the asteroids' regoliths is limited because the thermal parameter depends on both the thermal inertia and the rotational period. This is reflected in the large uncertainties in Table \ref{thermal}. Even with the uncertainties, however, some broad inferences can be made about which asteroids have surfaces which might be covered in rocky debris (thermal inertia $\sim50$ J/m$^2$/K/s$^{1/2}$) or sand ($\sim400$ J/m$^2$/K/s$^{1/2}$) versus which are bare rock ($\gtrsim$ 2500 J/m$^2$/K/s$^{1/2}$). Additionally, the Yarkovsky effect depends solely on the thermal parameter, so this characterization may help in learning more about the typical magnitude of the Yarkovsky effect, and thus its importance \citep{delbo}. 

This modeling may be most useful in the data it gives us on asteroid spin axes. Only a few hundred spin axes have been determined, as the commonly-used thermal models do not solve for them, and those that have been determined are biased toward larger asteroids \citep{krys}. While estimates have recently been made for the longitudes of many spin axes using Lowell Observatory lightcurves, this method requires asteroids which exhibit relatively large absolute brightness variation, and does not provide the asteroid's obliquity \citep{bowell}.  Characterization of asteroid obliquities and latitudes for smaller asteroids will help confirm spin axis distribution anisotropies and determine the relative importances of collisional, tidal, and thermal processes in altering spin directions, which is of some theoretical interest \citep{vokr}. Before using our results for these purposes it would be prudent to do follow-up testing of our results, including using more computational power and testing models which incorporate additional parameters such as non-spherical shapes.

As NEOWISE continues its three-year mission, the number of asteroids observed at multiple epochs will increase dramatically, and MCMC modeling using the SCREAM provides a method for characterizing these asteroids.\\

\section{Acknowledgments}

This work was supported by the NSF and began during an REU program at the University of California, Los Angeles.

SCK gratefully acknowledges the support of the Vagelos Challenge Award at the University of Pennsylvania.

This research has made use of the NASA/IPAC Infrared Science Archive, which is operated by the Jet Propulsion Laboratory, California Institute of Technology, under contract with the National Aeronautics and Space Administration.

This publication makes use of data products from the Wide-field Infrared Survey Explorer, which is a joint project of the University of California, Los Angeles, and the Jet Propulsion Laboratory/California Institute of Technology, and NEOWISE, which is a project of the Jet Propulsion Laboratory/California Institute of Technology. WISE and NEOWISE are funded by the National Aeronautics and Space Administration.

\bigskip

\begin{sidewaystable*}[p]
\captionsetup{labelfont=large}
\captionof{table}{\large Data}\label{data}
\begin{center}
\begin{tabular}{|l|r|r|r|r|r|r|r|r|r|r|r|r|r|r|}
\hline
Designation & MJD & $r$ (AU) & $\Delta$ (AU) & W1 flux (Jy) & W1 $\sigma$ & W2 flux & W2 $\sigma$ & W3 flux & W3 $\sigma$ & W4 flux & W4 $\sigma$ & Avg. SNR \\
\hline
\multirow{3}{*}{2014 HJ129} 
& 55208.4 & 1.134 & 0.582 & 3.1E-05 & 1.6E-05 & 2.92E-04 & 5.0E-05 & 7.49E-03 & 8.0E-04 & 1.03E-02 & 2.6E-03 & 5.3 \\ 
& 56704.8 & 1.184 & 0.623 & 3.2E-05 & 1.9E-05 & 2.91E-04 & 6.1E-05 & & & & & 3.2 \\ 
& 56773.2 & 1.227 & 0.619 & 1.9E-05 & 1.7E-05 & 2.51E-04 & 5.4E-05 & & & & & 2.9 \\ 
\hline 
\multirow{4}{*}{2009 UX17} 
& 55245.7 & 1.090 & 0.400 & 1.3E-05 & 1.6E-05 & 1.72E-04 & 3.9E-05 & 3.81E-03 & 4.9E-04 & 6.0E-03 & 2.3E-03 & 3.9 \\ 
& 55257.3 & 1.091 & 0.396 & 1.1E-05 & 1.6E-05 & 1.50E-04 & 3.9E-05 & 3.40E-03 & 4.7E-04 & 4.8E-03 & 2.3E-03 & 3.4 \\ 
& 56675.5 & 1.090 & 0.506 & 4.3E-05 & 2.0E-05 & 3.23E-04 & 5.9E-05 & & & & & 3.8 \\ 
& 56687.5 & 1.093 & 0.472 & 4.0E-05 & 1.7E-05 & 3.76E-04 & 5.8E-05 & & & & & 4.4 \\ 
\hline 
\multirow{4}{*}{2102 Tantalus} 
& 55365.7 & 1.658 & 1.209 & 2.9E-05 & 1.5E-05 & 1.09E-04 & 3.5E-05 & 5.80E-03 & 6.3E-04 & 1.02E-02 & 2.4E-03 & 4.6 \\ 
& 55390.0 & 1.675 & 1.224 & 3.0E-05 & 1.6E-05 & 9.4E-05 & 3.3E-05 & 5.51E-03 & 6.0E-04 & 9.9E-03 & 2.3E-03 & 4.5 \\ 
& 56663.6 & 1.128 & 0.595 & 5.28E-04 & 5.1E-05 & 3.99E-03 & 3.7E-04 & & & & & 10.6 \\ 
& 56816.5 & 1.154 & 0.676 & 4.15E-04 & 4.6E-05 & 2.99E-03 & 2.9E-04 & & & & & 9.8 \\ 
\hline 
\multirow{4}{*}{2000 PJ5} 
& 55380.8 & 1.195 & 0.629 & 4.1E-05 & 1.8E-05 & 3.39E-04 & 5.5E-05 & 9.73E-03 & 9.8E-04 & 1.56E-02 & 2.8E-03 & 6.0 \\ 
& 55388.5 & 1.186 & 0.563 & 5.2E-05 & 1.8E-05 & 4.28E-04 & 5.6E-05 & 1.17E-02 & 1.1E-03 & 1.75E-02 & 2.8E-03 & 6.7 \\ 
& 55406.2 & 1.149 & 0.487 & 6.9E-05 & 1.8E-05 & 6.02E-04 & 6.8E-05 & 1.48E-02 & 1.4E-03 & 2.05E-02 & 3.0E-03 & 7.5 \\ 
& 56833.2 & 1.172 & 0.579 & 7.5E-05 & 2.0E-05 & 5.23E-04 & 8.2E-05 & & & & & 5.0 \\ 
\hline 
\multirow{3}{*}{1998 SB15} 
& 55227.6 & 1.199 & 0.690 & 0.00E+00 & 1.4E-05 & 1.14E-04 & 4.2E-05 & 2.26E-03 & 4.4E-04 & 2.1E-03 & 2.5E-03 & 2.2 \\ 
& 55410.8 & 1.409 & 0.866 & 0.00E+00 & 8.3E-06 & 6.7E-05 & 3.7E-05 & 1.82E-03 & 4.0E-04 & 4.0E-03 & 2.3E-03 & 2.0 \\ 
& 56663.3 & 1.077 & 0.353 & 3.0E-05 & 1.7E-05 & 3.39E-04 & 6.0E-05 & & & & & 3.7 \\ 
\hline 
\multirow{4}{*}{2000 RJ34} 
& 55242.8 & 1.712 & 1.410 & 1.18E-04 & 2.1E-05 & 1.50E-03 & 2.6E-04 & 5.74E-02 & 5.4E-03 & 8.26E-02 & 8.4E-03 & 8.0 \\ 
& 55409.5 & 2.830 & 2.533 & 0.00E+00 & 1.5E-05 & 0.6E-05 & 3.7E-05 & 6.29E-03 & 7.0E-04 & 1.56E-02 & 2.7E-03 & 3.7 \\ 
& 56813.4 & 1.775 & 1.346 & 1.19E-04 & 2.6E-05 & 1.25E-03 & 1.4E-04 & & & & & 6.7 \\ 
& 56820.1 & 1.825 & 1.451 & 6.8E-05 & 2.2E-05 & 1.02E-03 & 1.1E-04 & & & & & 6.0 \\ 
\hline 
\multirow{2}{*}{2010 NG3} 
& 55386.6 & 1.443 & 0.947 & 3.0E-05 & 1.6E-05 & 2.77E-04 & 4.5E-05 & 9.99E-03 & 9.9E-04 & 1.53E-02 & 2.6E-03 & 6.0 \\ 
& 56823.1 & 1.159 & 0.546 & 3.00E-04 & 3.6E-05 & 2.91E-03 & 2.8E-04 & & & & & 9.3 \\ 
\hline 
\multirow{4}{*}{2003 LC5} 
& 55276.6 & 1.181 & 0.560 & 3.76E-04 & 4.0E-05 & 4.70E-03 & 4.4E-04 & 9.77E-02 & 9.1E-03 & 1.33E-01 & 1.3E-02 & 10.3 \\ 
& 55305.2 & 1.353 & 0.848 & 1.02E-04 & 2.0E-05 & 1.49E-03 & 1.5E-04 & 3.99E-02 & 3.7E-03 & 5.77E-02 & 6.0E-03 & 8.9 \\ 
& 55312.5 & 1.390 & 0.927 & 7.7E-05 & 1.8E-05 & 1.08E-03 & 1.1E-04 & 3.12E-02 & 2.9E-03 & 4.61E-02 & 5.0E-03 & 8.5 \\ 
& 56705.1 & 1.522 & 1.127 & 3.9E-05 & 2.2E-05 & 3.44E-04 & 6.2E-05 & & & & & 3.7 \\ 
\hline 
\multirow{3}{*}{2001 VS78} 
& 55406.3 & 2.337 & 2.089 & 1.4E-05 & 1.6E-05 & 0.9E-05 & 3.8E-05 & 2.07E-03 & 4.1E-04 & 4.3E-03 & 2.3E-03 & 2.0 \\ 
& 56737.9 & 1.258 & 0.780 & 2.72E-04 & 3.4E-05 & 1.79E-03 & 1.8E-04 & & & & & 9.0 \\ 
& 56744.8 & 1.270 & 0.775 & 3.10E-04 & 3.6E-05 & 1.96E-03 & 1.9E-04 & & & & & 9.4 \\ 
\hline 
\multirow{3}{*}{1999 GJ4} 
& 55214.6 & 2.375 & 2.105 & 0.8E-05 & 1.6E-05 & 0.00E+00 & 3.0E-05 & 2.36E-03 & 4.2E-04 & 7.0E-03 & 2.5E-03 & 2.2 \\ 
& 55341.8 & 1.744 & 1.432 & 6.9E-05 & 1.7E-05 & 1.39E-04 & 3.9E-05 & 7.91E-03 & 8.1E-04 & 1.18E-02 & 2.4E-03 & 5.6 \\ 
& 56660.8 & 1.454 & 1.038 & 1.22E-04 & 2.2E-05 & 6.62E-04 & 8.7E-05 & & & & & 6.6 \\ 
\hline 
\end{tabular}
\end{center}
A summary of the data used in our fits. The MJD given is the average modified Julian date of data points which were mid-averaged at that epoch. The RA and Dec are the RA and Dec of the asteroid as seen from Earth. $r$ and $\Delta$ are the distances between the asteroid and the Sun and Earth, respectively. Since WISE orbits the Earth, to the relevant accuracy $\Delta$ is also the distance between WISE and the asteroid. Fluxes and uncertainties in WISE's four bands are all in Janskys. A flux of 0 is a non-detection in that band. The average signal to noise ratio (SNR) is calculated by dividing the flux in each band by the uncertainty in that band, and averaging across that observation.
\end{sidewaystable*}

\begin{table*}{}
\captionsetup{labelfont=large}
\captionof{table}{\large Parameter Posterior Probability Distribution Characterization}\label{ppd}
\begin{center}
\begin{tabular}{|l|c|r|r|r|r|r|r|r|}
\hline
&  & Spin Axis & Spin Axis & Spin Axis & Thermal & Emissivity & Diameter & Albedo \\ 
&  & RA (J2000) & Dec (J2000) & Obliquity & Parameter &  & (meters) &  \\ \hline
\multirow{3}{*}{2014 HJ129} & $+1 \sigma$ & 153\degree & -49.4\degree & 56.2\degree & 2.74 & 0.980 & 536 & 0.053 \\ 
& Median & 140.\degree & -55.1\degree & 48.6\degree & 2.06 & 0.950 & 510. & 0.039 \\ 
& $-1 \sigma$ & 129\degree & -62.8\degree & 42.8\degree & 1.35 & 0.925 & 481 & 0.029 \\ 
\hline
\multirow{3}{*}{2009 UX17} & $+1 \sigma$ & 184.6\degree & 11.8\degree & 19.3\degree & 4.03 & 0.961 & 324 & 0.058 \\  
& Median & 178.3\degree & 4.0\degree & 12.6\degree & 3.39 & 0.935 & 306 & 0.043 \\ 
& $-1 \sigma$ & 171.9\degree & -4.7\degree & 9.3\degree & 2.98 & 0.910 & 289 & 0.032 \\ 
\hline
\multirow{3}{*}{2102 Tantalus} & $+1 \sigma$ & 116.2\degree & -29.2\degree & 81.2\degree & 1.23 & 0.966 & 1545 & 0.451 \\ 
& Median & 106.8\degree & -35.1\degree & 76.3\degree & 1.07 & 0.937 & 1503 & 0.342 \\ 
& $-1 \sigma$ & 96.4\degree & -41.6\degree & 70.5\degree & 0.89 & 0.911 & 1466 & 0.254 \\ 
\hline
\multirow{3}{*}{2000 PJ5} & $+1 \sigma$ & 198\degree & -38.8\degree & 99.2\degree & 0.99 & 0.973 & 777 & 0.411 \\  
& Median & 174\degree & -49.9\degree & 90.6\degree & 0.84 & 0.949 & 757 & 0.308 \\ 
& $-1 \sigma$ & 147\degree & -58.3\degree & 77.8\degree & 0.67 & 0.925 & 736 & 0.235 \\ 
\hline
\multirow{3}{*}{1998 SB15} & $+1 \sigma$ & 80.\degree & -7\degree & 50.\degree & 1.83 & 0.980 & 399 & 0.079 \\   
& Median & 66\degree & -20.\degree & 35\degree & 1.43 & 0.941 & 363 & 0.057 \\ 
& $-1 \sigma$ & 54\degree & -34\degree & 23\degree & 1.12 & 0.911 & 340. & 0.042 \\ 
\hline
\multirow{3}{*}{2000 RJ34} & $+1 \sigma$ & 222.8\degree & -4\degree & 23.4\degree & 0.29 & 0.973 & 3968 & 0.088 \\  
& Median & 214.7\degree & -17\degree & 12.4\degree & 0.19 & 0.943 & 3893 & 0.067 \\ 
& $-1 \sigma$ & 205.0\degree & -30.\degree & 7.0\degree & 0.08 & 0.915 & 3841 & 0.050 \\ 
\hline
\multirow{3}{*}{2010 NG3} & $+1 \sigma$ & 70.\degree & -22\degree & 54\degree & 1.01 & 0.964 & 1219 & 0.236 \\  
& Median & 50.\degree & -37\degree & 40.\degree & 0.81 & 0.928 & 1149 & 0.173 \\ 
& $-1 \sigma$ & 40.\degree & -49\degree & 30.\degree & 0.59 & 0.907 & 1095 & 0.129 \\ 
\hline
\multirow{3}{*}{2003 LC5} & $+1 \sigma$ & 17.1\degree & -48.9\degree & 41.6\degree & 2.81 & 0.977 & 1704 & 0.069 \\ 
& Median & 10.4\degree & -52.6\degree & 37.8\degree & 2.27 & 0.952 & 1678 & 0.052 \\ 
& $-1 \sigma$ & 5.1\degree & -56.0\degree & 34.3\degree & 1.73 & 0.925 & 1651 & 0.039 \\ 
\hline
\multirow{3}{*}{2001 VS78} & $+1 \sigma$ & 76\degree & 7\degree & 38.1\degree & 0.32 & 0.963 & 1663 & 0.56 \\ 
& Median & 51\degree & -18\degree & 19.4\degree & 0.18 & 0.930 & 1608 & 0.43 \\ 
& $-1 \sigma$ & 33\degree & -34\degree & 12.2\degree & 0.09 & 0.907 & 1579 & 0.32 \\ 
\hline
\multirow{3}{*}{1999 GJ4} & $+1 \sigma$ & 20.\degree & 17\degree & 45.8\degree & 0.61 & 0.980 & 1794 & 0.54 \\  
& Median & 7\degree & 1\degree & 34.8\degree & 0.38 & 0.946 & 1732 & 0.41 \\ 
& $-1 \sigma$ & 354\degree & -12\degree & 30.3\degree & 0.17 & 0.916 & 1673 & 0.30 \\ 
\hline
\end{tabular}
\end{center}
Posterior probability distribution characterization of the free parameters in the SCREAM applied to the 10 well-characterized asteroids (out of the 20 attempted characterizations) as well as the geometric visible albedo, a derived parameter. When deriving a posterior probability distribution for the  albedo, we needed to account for the uncertainty in external values we use to calculate it along with the diameter. We therefore built a PPD using values of $H$ drawn from a Gaussian of width 0.3 magnitudes centered at the value from JPL Horizons.  The $\pm 1 \sigma$ values are found by sorting the PPD for a parameter and taking the $50 \pm 34 \% $ values.
\end{table*}

\begin{table*}{}
\captionsetup{labelfont=large}
\captionof{table}{\large Thermal Inertia Estimates}\label{thermal}
\begin{center}
\renewcommand{\arraystretch}{1.2}
\begin{tabular}{|l|r|}
\hline
& Thermal Inertia \\
& (J/m$^2$/K/s$^{1/2}$) \\
\hline
2014 HJ129		&	$1390^{+720}_{-590}$ \\ \hline
2009 UX17	&	$2450^{+830}_{-870}$	 \\ \hline
2102 Tantalus		&		$670^{+230}_{-240}$	 \\ \hline
2000 PJ5	&		$540^{+200}_{-190}$	 \\ \hline
1998 SB15	&		$1000^{+420}_{-390}$	 \\ \hline
2000 RJ34	&		$120^{+97}_{-67}$	 \\ \hline
2010 NG3	&		$530^{+230}_{-210}$	 \\ \hline
2003 LC5	&		$1570^{+640}_{-620}$ \\ \hline
2001 VS78	&		$101^{+98}_{-55}$	 \\ \hline
1999 GJ4	&		$220^{+200}_{-130}$	 \\ \hline
\end{tabular}
\end{center}
Estimates of the thermal inertia of the fit asteroids. The thermal inertia depends on the thermal parameter, the albedo, the emissivity, and also the rotational frequency of the asteroid. We have PPDs of the first three of these from our fits, but not of the rotational frequency. We created PPDs for the thermal inertia by sampling the rotational frequency from a uniform distribution between (24 hours$)^{-1}$ and (2 hours$)^{-1}$, from the findings of \citet{pravec}. The thermal inertia values we give here are the median and $\pm 1 \sigma$ values of the derived PPDs.
\end{table*}

\begin{table*}{}
\captionsetup{labelfont=large}
\captionof{table}{\large Comparison of SCREAM and NEATM Models}\label{comparison}
\begin{center}
\renewcommand{\arraystretch}{1.2}
\begin{tabular}{|l|r|r|r|r|}
\hline
& SCREAM & NEATM & SCREAM & NEATM \\
& Diameter (m)& Diameter (m) & Albedo & Albedo \\
\hline
2014 HJ129		&	$510^{+26}_{-29}$	&	$519\pm190$		&	$0.039^{+0.014}_{-0.010}$	&	$0.031\pm0.027$ \\ \hline
2009 UX17	&	$306^{+18}_{-17}$		&	$309\pm10$		&	$0.043^{+0.015}_{-0.011}$	&	$0.042\pm0.008$ \\ \hline
2102 Tantalus		&		$1503^{+42}_{-37}$	&	$1810\pm214$	&	$0.342^{+0.109}_{-0.088}$	&	$0.214\pm0.084$ \\ \hline
2000 PJ5	&		$757^{+20}_{-21}$		&	$923\pm10$	&	$0.308^{+0.103}_{-0.073}$	&	$0.227\pm0.033$ \\ \hline
1998 SB15	&		$363^{+36}_{-23}$		&	$337\pm26$		&	$0.057^{+0.022}_{-0.015}$	&	$0.062\pm0.014$ \\ \hline
2000 RJ34	&		$3893^{+75}_{-52}$	&	$4330\pm100$	&	$0.067^{+0.021}_{-0.017}$	&	$0.067\pm0.011$ \\ \hline
2010 NG3	&		$1149^{+70}_{-54}$	&	$1520\pm40$	&	$0.173^{+0.063}_{-0.044}$	&	$0.100\pm0.022$ \\ \hline
2003 LC5	&		$1667^{+26}_{-27}$	&	$1678\pm10$	&	$0.052^{+0.017}_{-0.013}$	&	$0.048\pm0.048$ \\ \hline
2001 VS78	&		$1608^{+55}_{-29}$	&	$2000\pm400$	&	$0.425^{+0.137}_{-0.107}$	&	$0.240\pm0.131$ \\ \hline
1999 GJ4	&		$1732^{+62}_{-59}$	&	$1641\pm53$	&	$0.406^{+0.130}_{-0.102}$	&	$0.453\pm0.087$ \\ \hline
\end{tabular}
\end{center}
A comparison between the modeled diameters and albedos from the SCREAM and those from the NEATM, which were found in \citet{mainzerc} and \citet{mainzer4}. The albedos are the geometric visible albedos. Note that best-fit absolute magnitude estimates from the JPL Horizons system may change over time as more data becomes available, so the values used in earlier works and in this paper may be different. This would in turn affect the relationship between the diameter and the albedo.
\end{table*}

%\begin{figure*}[h]
%\centering
%\includegraphics[width=0.7\textwidth]{FIG8.jpg}
%\caption{A 2D histogram of the posterior probability distributions of the spin axis orientation RA and Dec (J2000) from the MCMC of the SCREAM applied to the asteroid 2014 HJ129, projected onto the sphere of the sky. The azimuthal viewing angle has been rotated so that the camera is facing the bin of highest probability. The elevation viewing angle is zero. \label{2014HJ129JointHist}}
%\end{figure*}

%\begin{figure*}[h]
%\centering
%\includegraphics[width=0.7\textwidth]{FIG11.jpg}
%\caption{A 2D histogram of the posterior probability distributions of the spin axis orientation RA and Dec (J2000) from the MCMC of the SCREAM applied to the asteroid 1975 YA, projected onto the sphere of the sky. The azimuthal viewing angle has been rotated so that the camera is facing the bin of highest probability. The elevation viewing angle is zero. \label{1975YAJointHist}}
%\end{figure*}

\end{document}